\documentclass[12pt]{article}
\usepackage{siunitx}
\usepackage{graphicx}
\usepackage[utf8]{inputenc}
\usepackage{graphicx,xcolor,setspace,geometry}
\usepackage{amsmath,setspace}
\usepackage{amssymb}
\usepackage{mathtools}
\usepackage{multirow}
\usepackage{booktabs}
\usepackage{threeparttable}
\usepackage{adjustbox}
\usepackage{nicefrac}
\usepackage{natbib}
\usepackage{multirow}
\usepackage{color}
\usepackage{tikz}
\usetikzlibrary{shapes}
\usetikzlibrary{plotmarks}
\usetikzlibrary{tikzmark,matrix,calc}
\usetikzlibrary{arrows.meta,calc,decorations.markings,math,arrows.meta}
\usepackage{caption}
\usepackage{hyperref}

\title{Continuous-time state-space modelling of the hot hand in basketball}

\author{Sina Mews$^{1,}\footnote{Corresponding author; email: \texttt{sina.mews@uni-bielefeld.de}.}$ and Marius Ötting$^{1}$ \\
 \\
$^{1}$Bielefeld University, Germany}

\date{}

\begin{document}

\begin{spacing}{1.25}
    \maketitle
\end{spacing}


\begin{spacing}{1.5}

\begin{abstract}
We investigate the hot hand phenomenon using data on 110,513 free throws taken in the National Basketball Association (NBA). 
As free throws occur at unevenly spaced time points within a game, we consider a state-space model formulated in continuous time to investigate serial dependence in players' success probabilities.
In particular, the underlying state process can be interpreted as a player's (latent) varying form and is modelled using the Ornstein-Uhlenbeck process.
Our results support the existence of the hot hand, but the magnitude of the estimated effect is rather small. 
\end{abstract}


\section{Introduction}
\label{s:intro}

In several areas of society, it remains an open question whether a ``hot hand'' effect exists, according to which humans may temporarily enter a state during which they perform better than on average.
While this concept may occur in different fields, such as among hedge fund managers and artists \citep{jagannathan2010hot,liu2018hot}, it is most prominent in sports. 
Sports commentators and fans --- especially in basketball --- often refer to players as having a ``hot hand'', and being ``on fire'' or ``in the zone'' when they show a (successful) streak in performance. 
In the academic literature, the hot hand has gained great interest since the seminal paper by \citet{gilovich1985hot}, who investigated a potential hot hand effect in basketball. 
They found no evidence for its existence and attributed the hot hand to a cognitive illusion, much to the disapproval of many athletes and fans. 
Still, the results provided by	\citet{gilovich1985hot} have often been used as a primary example for showing that humans over-interpret patterns of success and failure in random sequences (see, e.g., \citealp{thaler2009nudge,kahneman2011thinking}). 

During the last decades, many studies have attempted to replicate or refute the results of \citet{gilovich1985hot}, analysing sports such as, for example, volleyball, baseball, golf, and especially basketball.
\citet{bar2006twenty} provide an overview of 24 studies on the hot hand, 11 of which were in favour of the hot hand phenomenon. 
Several more recent studies, often based on large data sets, also provide evidence for the existence of the hot hand (see, e.g., \citealp{raab2012hot,green2017hot,miller2016surprised,chang2019predictive}).
Notably, \citet{miller2016surprised} show that the original study from \citet{gilovich1985hot} suffers from a selection bias.
Using the same data as in the original study by \citet{gilovich1985hot}, \citet{miller2016surprised} account for that bias, and their results do reveal a hot hand effect. 
However, there are also recent studies which provide mixed results (see, e.g., \citealp{wetzels2016bayesian}) or which do not find evidence for the hot hand, such as \citet{morgulev2020searching}. 
Thus, more than 30 years after the study of \citet{gilovich1985hot}, the existence of the hot hand remains highly disputed. 

Moreover, the literature does not provide a universally accepted definition of the hot hand effect. 
While some studies regard it as serial correlation in \textit{outcomes} (see, e.g., \citealp{gilovich1985hot,dorsey2004bowlers,miller2016surprised}), others consider it as serial correlation in \textit{success probabilities} (see, e.g., \citealp{albert1993statistical,wetzels2016bayesian,otting2018hot}).
The latter definition translates into a latent (state) process underlying the observed performance --- intuitively speaking, a measure for a player's form --- which can be elevated without the player necessarily being successful in every attempt. 
In our analysis, we follow this approach and hence consider state-space models (SSMs) to investigate the hot hand effect in basketball.
Specifically, we analyse free throws from more than 9,000 games played in the National Basketball Association (NBA), totalling in 110,513 observations. 
In contrast, \citet{gilovich1985hot} use data on 2,600 attempts in their controlled shooting experiment.

Free throws in basketball, or similar events in sports with game clocks, occur at unevenly spaced time points.
These varying time lengths between consecutive attempts may affect inference on the hot hand effect if the model formulation does not account for the temporal irregularity of the observations. 
As an illustrative example, consider an irregular sequence of throws with intervals ranging from, say, two seconds to 15 minutes. 
For intervals between attempts that are fairly short (such as a few seconds), a player will most likely be able to retain his form from the last shot. 
On the other hand, if several minutes elapse before a player takes his next shot, it becomes less likely that he is able to retain his form from the last attempt. 
However, we found that existing studies on the hot hand do not account for different interval lengths between attempts. 
In particular, studies investigating serial correlation in success probabilities usually consider discrete-time models that require the data to follow a regular sampling scheme and thus, cannot (directly) be applied to irregularly sampled data.
In our contribution, we overcome this limitation by formulating our model in continuous time to explicitly account for irregular time intervals between free throws in basketball. 
Specifically, we consider a stochastic differential equation (SDE) as latent state process, namely the Ornstein-Uhlenbeck (OU) process, which represents the underlying form of a player fluctuating continuously around his average performance. 

In the following, Section \ref{s:data} presents our data set and covers some descriptive statistics. 
Subsequently, in Section \ref{s:model}, the continuous-time state-space model (SSM) formulation for the analysis of the hot hand effect is introduced, while its results are presented in Section \ref{s:results}.
We conclude our paper with a discussion in Section \ref{s:discussion}.

\section{Data}
\label{s:data}

We extracted data on all basketball games played in the NBA between October 2012 and June 2019 from \url{https://www.basketball-reference.com/}, covering both regular seasons and playoff games. 
For our analysis, we consider data only on free throw attempts as these constitute a highly standardised setting without any interaction between players, which is usually hard to account for when modelling field goals in basketball. 
We further included all players who took at least 2,000 free throws in the period considered, totalling in 110,513 free throws from 44 players. 
For each player, we included only those games in which he attempted at least four free throws to ensure that throws did not only follow successively (as players receive up to three free throws if they are fouled). 
A single sequence of free throw attempts consists of all throws taken by one player in a given game, totalling in 15,075 throwing sequences with a median number of 6 free throws per game (min: 4; max: 39). 

As free throws occur irregularly within a basketball game, the information on whether an attempt was successful needs to be supplemented by its time point $t_k, k = 1,\ldots,T$, where $0 \leq t_1 \leq t_2 \leq \ldots \leq t_T$, corresponding to the time already played (in minutes) as indicated by the game clock. 
For each player $p$ in his $n$-th game, we thus consider an irregular sequence of binary variables $Y_{t_1}^{p,n}, Y_{t_2}^{p,n}, \ldots, Y_{t_{T_{p,n}}}^{p,n}$, with 
$$ 
Y_{t_k}^{p,n} = \begin{cases}
    1 & \text{if free throw attempt at time $t_k$ is successful;} \\
    0 & \text{otherwise}.
\end{cases} 
$$
In our sample, the proportion of successful free throw attempts is obtained as 0.784. 
However, there is considerable heterogeneity in the players' throwing success as the corresponding empirical proportions range from 0.451 (Andre Drummond) to 0.906 (Stephen Curry).
Players can receive up to three free throws (depending on the foul) in the NBA, which are then thrown in quick succession, and the proportion of successful free throws differs substantially between the three attempts, with 0.769, 0.8, and 0.883 obtained for the first, second, and third free throw, respectively. 
To account for the position of the throw in a player's set of (at most) three free throws, we hence include the dummy variables \textit{ft2} and \textit{ft3} in our analysis. 
In our sample, 54.5\% of all free throws correspond to the first, 43.7\% to the second, and only 1.8\% to the third attempt in a set (cf.\ Table~\ref{tab:descr}). 
Furthermore, as the outcome of a free throw is likely affected by intermediate information on the game --- such as a close game leading to pressure situations --- we consider several further covariates, which were also used in previous studies (see, e.g., \citealp{toma2017missed,morgulev2020searching}). 
Specifically, we consider the current score difference (\textit{scorediff}), a home dummy (\textit{home}), and a dummy indicating whether the free throw occurred in the last 30 seconds of the quarter (\textit{last30}). 
Corresponding summary statistics are shown in Table~\ref{tab:descr}.

\begin{table}[!b] \centering 
  \caption{Descriptive statistics of the covariates.} 
  \label{tab:descr} 
\begin{tabular}{@{\extracolsep{5pt}}lccccc} 
\\[-1.8ex]\hline 
\hline \\[-1.8ex] 
 &  \multicolumn{1}{c}{mean} & \multicolumn{1}{c}{st.\ dev.} & \multicolumn{1}{c}{min.} & \multicolumn{1}{c}{max.} \\ 
\hline \\[-1.8ex] 
\textit{scorediff} & 0.576 & 9.860 & $-$45 & 49 \\ 
\textit{home} &  0.514 & -- & 0 & 1 \\ 
\textit{last30} &  0.093 & -- & 0 & 1 \\ 
\textit{ft2} &  0.437 & -- & 0 & 1 \\ 
\textit{ft3} &  0.018 & -- & 0 & 1 \\ 
\hline \\[-1.8ex] 
\end{tabular} 
\end{table}

In Table~\ref{tab:data}, example throwing sequences used in our analysis are shown for free throws taken by LeBron James in five NBA games.
These throwing sequences illustrate that free throw attempts often appear in clusters of two or three attempts at the same time (depending on the foul), followed by a time period without any free throws.
Therefore, it is important to take into account the different lengths of the time intervals between consecutive attempts as the time elapsed between attempts affects a player's underlying form.

\begin{table}[!t]
\centering
\caption{Throwing sequences of LeBron James.}
\label{tab:data}
\scalebox{0.75}{
\begin{tabular}{llllllllllllll}
  \hline
  \multicolumn{13}{c}{Miami Heat @ Houston Rockets, November 12, 2012} \\
$y_{t_k}^{\text{James},1}$ & 0 & 1 & 0 & 1 & 1 & 0 & 1 & 1 & - & - & - & - & - \\ 
  $t_k$ (in min.) & 11.01 & 11.01 & 23.48 & 23.48 & 46.68 & 46.68 & 47.09 & 47.09 & - & - & - & - & - \\ 
  \hline
  \multicolumn{13}{c}{Miami Heat @ Los Angeles Clippers, November 14, 2012} \\
  $y_{t_k}^{\text{James},2}$ & 0 & 1 & 0 & 0 & 1 & 1 & 1 & 1 & 1 & - & - & - & - \\ 
  $t_k$ (in min.) & 9.47 & 23.08 & 23.08 & 24.73 & 36.95 & 36.95 & 41 & 41 & 42.77 & - & - & - & - \\ 
  \hline
  \multicolumn{13}{c}{Milwaukee Bucks @ Miami Heat, November 21, 2012} \\
  $y_{t_k}^{\text{James},3}$ & 0 & 1 & 0 & 1 & 0 & 1 & 1 & 0 & - & - & - & - & - \\ 
  $t_k$ (in min.) & 7.5 & 7.5 & 10.02 & 10.02 & 35.62 & 35.62 & 42.68 & 42.68 & - & - & - & - & - \\ 
  \hline
  \multicolumn{13}{c}{Cleveland Cavaliers @ Miami Heat, November 24, 2012} \\
  $y_{t_k}^{\text{James},4}$ & 1 & 1 & 1 & 0 & 1 & 1 & 1 & 1 & 1 & - & - & - & - \\ 
  $t_k$ (in min.) & 11.62 & 21.07 & 21.07 & 21.38 & 21.38 & 31.95 & 31.95 & 32.68 & 32.68 & - & - & - & - \\ 
  \hline
  \multicolumn{13}{c}{Los Angeles Lakers @ Miami Heat, January 23, 2014} \\
  $y_{t_k}^{\text{James},5}$ & 0 & 1 & 0 & 1 & 1 & 0 & 0 & 1 & 1 & 0 & 1 & 0 & 1 \\ 
  $t_k$ (in min.) & 9.28 & 9.28 & 20.53 & 25.97 & 25.97 & 31.57 & 31.57 & 33.43 & 33.43 & 42.1 & 42.1 & 47.62 & 47.62 \\ 
   \hline
\end{tabular}}
\end{table}

\section{Continuous-time modelling of the hot hand}
\label{s:model}

\subsection{State-space model specification}

\begin{figure}[!htb]
    \centering
\begin{tikzpicture}[scale=0.75, transform shape]
	\node[circle,draw=black, fill=gray!5, inner sep=0pt, minimum size=50pt] (A) at (2, -5) {$S_{t_1}$};
	\node[circle,draw=black, fill=gray!5, inner sep=0pt, minimum size=50pt] (B) at (4.4, -5) {$S_{t_2}$};
	\node[circle,draw=black, fill=gray!5, inner sep=0pt, minimum size=50pt] (C) at (7.5, -5) {$S_{t_3}$};
	\node[circle,draw=black, fill=gray!5, inner sep=0pt, minimum size=50pt] (C1) at (10, -5) {...};
	\node[circle,draw=black, fill=gray!5, inner sep=0pt, minimum size=50pt] (Y1) at (2, -2.5) {$Y_{t_1}$};
	\node[circle,draw=black, fill=gray!5, inner sep=0pt, minimum size=50pt] (Y2) at (4.4, -2.5) {$Y_{t_2}$};
	\node[circle,draw=black, fill=gray!5, inner sep=0pt, minimum size=50pt] (Y3) at (7.5, -2.5) {$Y_{t_3}$};
	\node[text width=5cm,font=\itshape] at (14, -2.5) 
    {throwing success (observed)};
    \node[text width=5cm,font=\itshape] at (14, -5) 
    {player's form (hidden)};
	\draw[-{Latex[scale=2]}] (A)--(B);
	\draw[-{Latex[scale=2]}] (B)--(C);
	\draw[-{Latex[scale=2]}] (C)--(C1);
	\draw[-{Latex[scale=2]}] (A)--(Y1);
	\draw[-{Latex[scale=2]}] (B)--(Y2);
	\draw[-{Latex[scale=2]}] (C)--(Y3);
\end{tikzpicture}
    \caption{Dependence structure of our SSM: the throwing success $Y_{t_k}$ is assumed to be driven by the underlying (latent) form of a player. To explicitly account for the irregular time intervals between observations, we formulate our model in continuous time.}
    \label{fig:hmm_bild}
\end{figure}
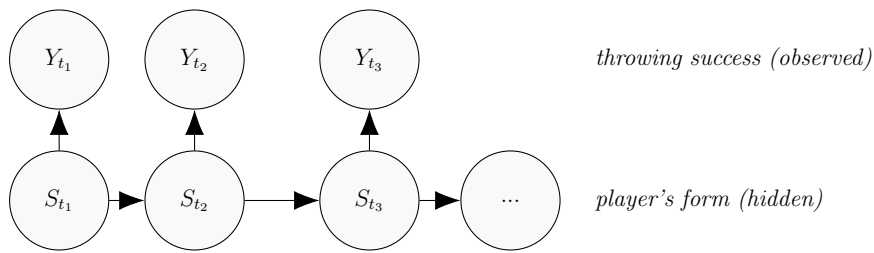

Following the idea that the throwing success depends on a player's current (latent) form \citep[see, e.g.,][]{albert1993statistical,wetzels2016bayesian,otting2018hot}, we model the observed free throw attempts using a state-space model formulation as represented in Figure~\ref{fig:hmm_bild}. 
The observation process corresponds to the binary sequence of a player's throwing success, while the state process can be interpreted as a player's underlying form (or ``hotness'').
We further include the covariates introduced in Section~\ref{s:data} in the model, which possibly affect a player's throwing success. 
In particular, we model the binary response of throwing success $Y_{t_k}^{p,n}$ using a Bernoulli distribution with the associated success probability $\pi_{t_k}^{p,n}$ being a function of the player's current state $S_{t_k}^{p,n}$ and the covariates. 
Dropping the superscripts $p$ and $n$ for notational simplicity from now on, we thus have
\begin{equation}
\label{eq:observations}
\begin{split}
    Y_{t_k} \sim \text{Bern}(\pi_{t_k}), \quad
    \text{logit}(\pi_{t_k}) = S_{t_k} &+ \beta_{0,p} + \beta_1 \textit{home} + \beta_2 \textit{scorediff} \\ &+ \beta_3 \textit{last30} + \beta_4 \textit{ft2} + \beta_5 \textit{ft3},
\end{split}
\end{equation}
where $\beta_{0,p}$ is a player-specific intercept to account for differences between players' average throwing success. 
To address the temporal irregularity of the free throw attempts, we formulate the stochastic process $\{S_t\}_{t \geq 0}$ in continuous time.
Furthermore, we require the state process to be continuous-valued to allow for gradual changes in a player's form, rather than assuming a finite number of discrete states \citep[e.g.\ three states interpreted as cold vs.\ normal vs.\ hot; cf.][]{wetzels2016bayesian, green2017hot}.
In addition, the state process ought to be stationary such that in the long-run a player returns to his average form.
A natural candidate for a corresponding stationary, continuous-valued and continuous-time process is the OU process, which is described by the following SDE:
\begin{equation}
    d S_t = \theta (\mu - S_t) dt + \sigma d B_t, \quad S_0=s_0,
\label{eq:OUprocess}
\end{equation}
where $\theta > 0$ is the drift term indicating the strength of reversion to the long-term mean $\mu \in \mathbb{R}$, while $\sigma > 0$ is the diffusion parameter controlling the strength of fluctuations, and $B_t$ denotes the Brownian motion.
We further specify $\mu = 0$ such that the state process fluctuates around a player's average form, given the current covariate values. 
Specifically, positive values of the state process indicate higher success probabilities, whereas negative values indicate decreased throwing success given the player's average ability and the current game characteristics.

As shown in Figure~\ref{fig:hmm_bild}, we model the hot hand effect as serial correlation in success probabilities as induced by the state process:
while the observed free throw attempts are conditionally independent, given the underlying states, the unobserved state process induces correlation in the observation process.
Regarding the hot hand effect, the drift parameter $\theta$ of the OU process is thus of main interest as it governs the speed of reversion (to the average form).
The smaller $\theta$, the longer it takes for the OU process to return to its mean
and hence the higher the serial correlation.
To assess whether a model including serial dependence (i.e.\ an SSM) is actually needed to describe the structure in the data, we additionally fit a benchmark model without the underlying state variable $S_{t_k}$ in Equation (\ref{eq:observations}).
Consequently, the benchmark model corresponds to the absence of any hot hand effect, i.e.\ a standard logistic regression model.

\subsection{Statistical inference}
\label{s:statInf}

The likelihood of the continuous-time SSM given by Equations (\ref{eq:observations}) and (\ref{eq:OUprocess}) involves integration over all possible realisations of the continuous-valued state $S_{t_k}$, at each observation time $t_1, t_2, \ldots, t_T$. 
For simplicity of notation, let the integer $\tau = 1, 2, \ldots, T$ denote the \textit{number} of the observation in the time series, such that $Y_{t_\tau}$ shortens to $Y_\tau$ and $S_{t_\tau}$ shortens to $S_\tau$. 
Further, $t_\tau$ represents the \textit{time} at which the observation $\tau$ was collected. 
Then the likelihood of a single throwing sequence $y_1, \ldots, y_T$ is given by
\begin{equation}
\begin{split}
    \mathcal{L}_T &= \int \ldots \int \text{Pr}(y_1, \ldots, y_T, s_1, \ldots, s_T) ds_T \ldots ds_1 \\
    &= \int \ldots \int \text{Pr}(s_1) \text{Pr}(y_1|s_1) \prod_{\tau=2}^T \text{Pr}(s_\tau | s_{\tau-1}) \text{Pr}(y_\tau|s_\tau) ds_T \ldots ds_1, \\
\label{eq:llk}
\end{split}
\end{equation}
where we assume that each player starts a game in his stationary distribution $S_1 \sim \mathcal{N} \left(0, \frac{\sigma^2}{2\theta} \right)$, i.e.\ the stationary distribution of the OU process.
Further, we assume $Y_\tau$ to be Bernoulli distributed with corresponding state-dependent probabilities $\text{Pr}(y_\tau|s_\tau)$ additionally depending on the current covariate values (cf.\ Equation~(\ref{eq:observations})),
while the probabilities of transitioning between the states $\text{Pr}(s_\tau | s_{\tau-1})$ are normally distributed as determined by the conditional distribution of the OU process:
\begin{equation}
    S_{\tau} | S_{\tau-1} = s \sim \mathcal{N}\left( \text{e}^{-\theta \Delta_\tau} s, \quad 
    \frac{\sigma^2}{2\theta} \bigl(1- \text{e}^{-2\theta \Delta_\tau}\bigr) \right),
\label{eq:condDistOU}
\end{equation}
where $\Delta_\tau  = t_{\tau} - t_{\tau-1}$ denotes the time difference between consecutive observations.

Due to the $T$ integrals in Equation (\ref{eq:llk}), the likelihood calculation is intractable. 
To render its evaluation feasible, we approximate the multiple integral by finely discretising the continuous-valued state space as first suggested by \citet{kitagawa1987}.
The discretisation of the state space can effectively be seen as a reframing of the model as a continuous-time hidden Markov model (HMM) with a large but finite number of states, enabling us to apply the corresponding efficient machinery. 
In particular, we use the forward algorithm to calculate the likelihood, defining the possible range of state values as $[-2, 2]$, which we divide into 100 intervals.
For details on the approximation of the likelihood via state discretisation, see \citet{otting2018hot} for discrete-time and \citet{mews2020} for continuous-time SSMs.

Assuming single throwing sequences of players to be mutually independent, conditional on the model parameters, the likelihood over all games and players is simply calculated as the product of the individual likelihoods.
The model parameters, i.e.\ the drift parameter and the diffusion coefficient of the OU process as well as the regression coefficients, are then estimated by numerically maximising the (approximate) joint likelihood. 
The resulting parameter estimators are unbiased and consistent --- corresponding simulation experiments are shown in \citet{mews2020}.

\section{Results}
\label{s:results}

\begin{table}[!b] \centering
\caption{\label{tab:results} Parameter estimates with 95\% confidence intervals.}
\medskip
\begin{tabular}{llcc}
    \\[-1.8ex]\hline 
\hline \\[-1.8ex] 
    \multicolumn{2}{l}{parameter} & estimate & 95\% CI \\
    \hline \\[-1.8ex] 
    $\theta$ & (drift) & 0.042  & [0.016; 0.109]  \\ 
    $\sigma$ & (diffusion) & 0.101  & [0.055; 0.185]  \\
    $\beta_1$ & (\textit{home}) & 0.023  & [-0.009; 0.055]  \\ 
    $\beta_2$ & (\textit{scorediff}) & 0.030  & [0.011; 0.048]  \\ 
    $\beta_3$ & (\textit{last30}) & 0.003  & [-0.051; 0.058]  \\ 
    $\beta_4$ & (\textit{ft2}) & 0.223  & [0.192; 0.254]  \\ 
    $\beta_5$ & (\textit{ft3}) & 0.421  & [0.279; 0.563]  \\ 
    \hline \\[-1.8ex] 
\end{tabular}
\end{table}

\begin{figure}[!t] \centering
\includegraphics[width=0.9\textwidth]{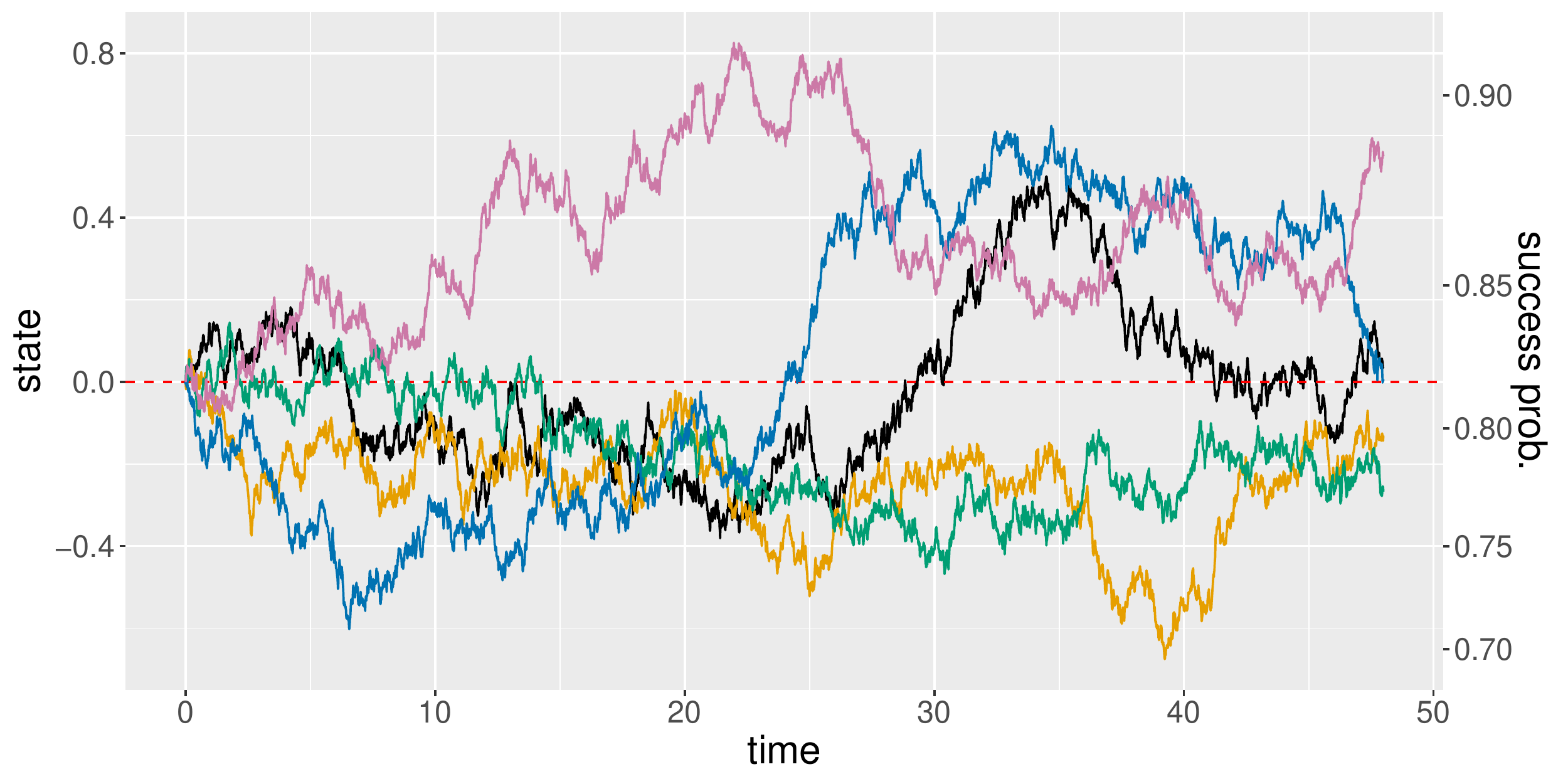}
\caption{\label{fig:OUresults} Simulation of possible state trajectories for the length of an NBA game based on the estimated parameters of the OU process. The red dashed line indicates the intercept (here: the median throwing success over all players), around which the processes fluctuate. The right y-axis shows the success probabilities resulting from the current state (left y-axis), given that the explanatory variables equal 0. The graphs were obtained by application of the Euler-Maruyama scheme with initial value 0 and step length 0.01.}
\end{figure}

According to the information criteria AIC and BIC, the continuous-time model formulation including a potential hot hand effect is clearly favoured over the benchmark model without any underlying state process ($\Delta$AIC = 61.20, $\Delta$BIC = 41.97). 
The parameter estimates of the OU process, which represents the underlying form of a player, as well as the estimated regression coefficients are shown in Table~\ref{tab:results}.
In particular, the estimate for the drift parameter $\theta$ of the OU process is fairly small, thus indicating serial correlation in the state process over time. 
However, when assessing the magnitude of the hot hand effect, we also observe a fairly small estimate for the diffusion coefficient $\sigma$. 
The corresponding stationary distribution is thus estimated as $\mathcal{N} \left(0,  0.348^2 \right)$, indicating a rather small range of the state process, which becomes apparent also when simulating state trajectories based on the parameter estimates of the OU process (cf.\ Figure~\ref{fig:OUresults}).
Still, the associated success probabilities, given that all covariate values are fixed to zero, vary considerably during the time of an NBA game (cf.\ right y-axis of Figure~\ref{fig:OUresults}).
While the state process and hence the resulting success probabilities slowly fluctuate around the average throwing success (given the covariates), the simulated state trajectories reflect the temporal persistence of the players' underlying form.
Thus, our results suggest that players can temporarily enter a state in which their success probability is considerably higher than their average performance, which provides evidence for a hot hand effect.

Regarding the estimated regression coefficients, the player-specific intercepts $\hat{\beta}_{0,p}$ range from -0.311 to 2.192 (on the logit scale), reflecting the heterogeneity in players' throwing success.
The estimates for $\beta_1$ to $\beta_5$ are displayed in Table \ref{tab:results} together with their 95\% confidence intervals. 
The chance of making a free throw is slightly increased if the game is played at home ($\beta_1$) or if a free throw occurs in the last 30 seconds of a quarter ($\beta_3$), but both corresponding confidence intervals include the zero. 
In contrast, the confidence interval for the score difference ($\beta_2$) does not include the zero and its effect is positive but small, indicating that the higher the lead, the higher is, on average, the chance to make a free throw. 
The position of the throw, i.e.\ whether it is the first, second ($\beta_4$), or third ($\beta_5$) attempt in a row, has the largest effect of all covariates considered:
compared to the first free throw, the chance of a hit is considerably increased if it is the second or, in particular, the third attempt, which was already indicated by the descriptive analysis presented in Section \ref{s:data}. 
However, this strong effect on the success probabilities is probably caused by the fact that three free throws in a row are only awarded if a player is fouled while shooting a three-point field goal, which, in turn, is more often attempted by players who regularly perform well at free throws. 

\begin{figure}[!h tb] \centering
\includegraphics[width=0.95\textwidth]{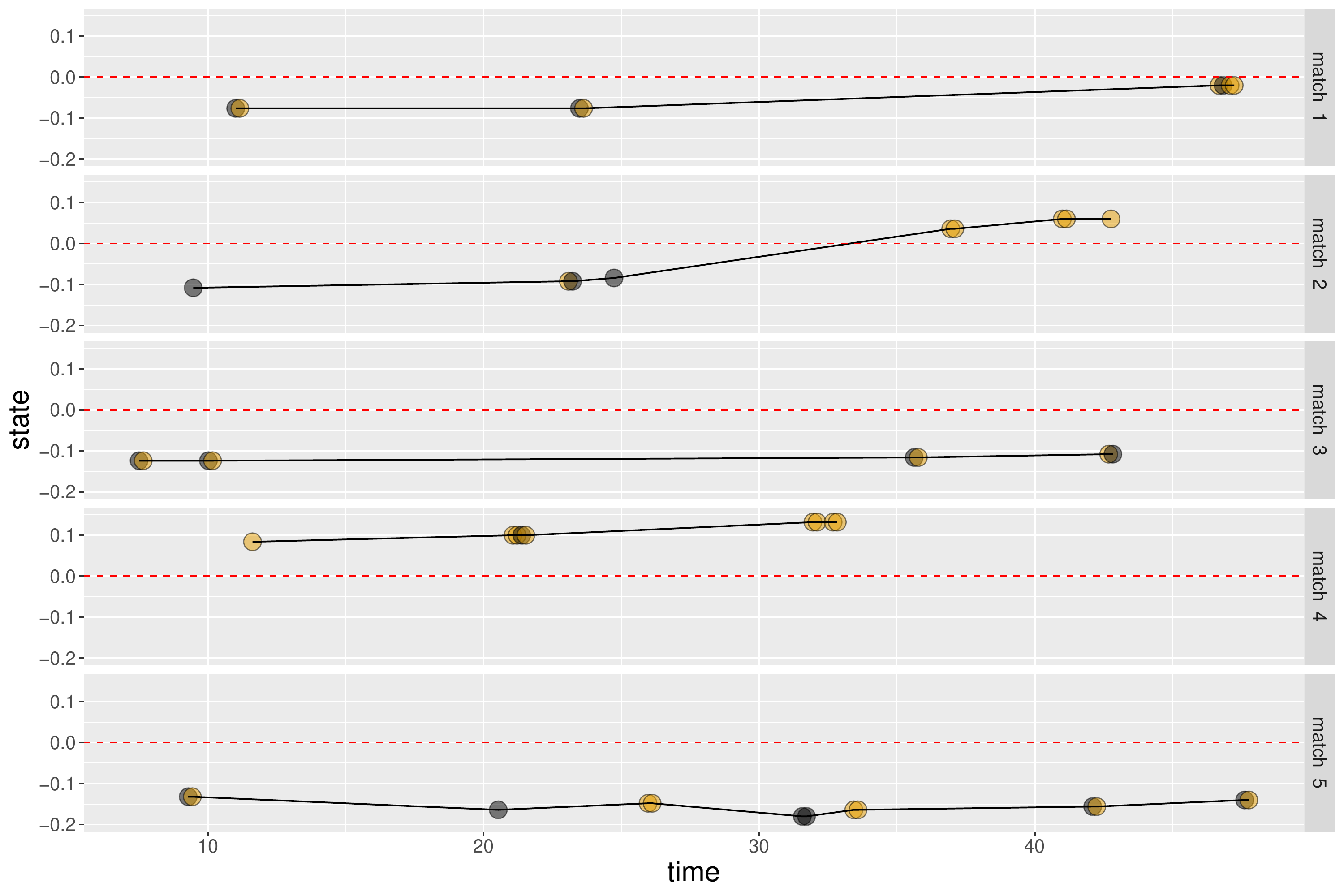}
\caption{\label{fig:decStates} Decoded states underlying the throwing sequences of LeBron James shown in Table~\ref{tab:data}. Successful free throws are shown in yellow, missed shots in black.}
\end{figure}

To further investigate how the hot hand may evolve during a game, we compute the most likely state sequences, corresponding to the underlying form of a player. 
Specifically, we seek
$$ 
(s_{1}^*,\ldots,s_{T}^*) = \underset{s_{1},\ldots,s_{T}}{\operatorname{argmax}} \; \Pr ( s_{1},\ldots,s_{T} | y_{1},\ldots, y_{T} ), 
$$
where $s_{1}^*,\ldots,s_{T}^*$ denotes the most likely state sequence given the observations. 
As we transferred our continuous-time SSM to an HMM framework by finely discretising the state space (cf.\ Section \ref{s:statInf}), we can use the Viterbi algorithm to calculate such sequences at low computational cost \citep{zucchini2016hidden}.
Figure~\ref{fig:decStates} shows the most likely states underlying the throwing sequences presented in Table~\ref{tab:data}. 
While the decoded state processes fluctuate around zero (i.e.\ a player's average throwing success), the state values vary slightly over the time of an NBA game.
Over all players, the decoded states range from -0.42 to 0.46, again indicating that the hot hand effect as modelled by the state process is rather small.

The decoded state sequences in Figure \ref{fig:decStates} further allow to illustrate the advantages and the main idea of our continuous-time modelling approach. 
For example, consider the throwing sequence in the second match shown, where LeBron James only made a single free throw of his first four attempts. 
The decoded state at throw number 3 is -0.092 (cf.\ Figure~\ref{fig:decStates}) and the time passed between throw number 3 and 4 is 1.65 minutes (cf.\ Table~\ref{tab:data}). 
Thus, the value of the state process at throw number 4 is drawn from a normal distribution, given the decoded state of the previous attempt, with mean $\text{e}^{-0.042 \cdot 1.65} (-0.092) = -0.086$ and variance $\frac{0.101^2}{2\cdot 0.042} (1 - \text{e}^{-2\cdot 0.042 \cdot 1.65}) = 0.016$ (cf.\ Equation~(\ref{eq:condDistOU})).
Accordingly, the value of the state process for throw number 5 is drawn from a normal distribution with mean $-0.050$ and variance $0.078$, conditional on the decoded state of -0.084 at throw number 4 and a relatively long time interval of 12.22 minutes. 
As highlighted by these example calculations, the conditional distribution of the state process takes into account the interval length between consecutive attempts:
the more time elapses, the higher the variance in the state process and hence, the less likely is a player to retain his form, with a tendency to return to his average performance.

\section{Discussion}
\label{s:discussion}

In our analysis of the hot hand, we used SSMs formulated in continuous time to model throwing success in basketball. 
Focusing on free throws taken in the NBA, our results provide evidence for a hot hand effect as the underlying state process exhibits some persistence over time. 
In particular, the model including a hot hand effect is preferred over the benchmark model without any underlying state process by information criteria. 
Although we provide evidence for the existence of a hot hand, the magnitude of the hot hand effect is rather small as the underlying success probabilities are only elevated by a few percentage points (cf.\ Figures~\ref{fig:OUresults} and \ref{fig:decStates}). 

A minor drawback of the analysis arises from the fact that there is no universally accepted definition of the hot hand. 
In our setting, we use the OU process to model players' continuously varying form and it is thus not clear which values of the drift parameter $\theta$ correspond to the existence of the hot hand. 
While lower values of $\theta$ refer to a slower reversion of a player's form to his average performance, a further quantification of the magnitude of the hot hand effect is not possible.
In particular, a comparison of our results to other studies on the hot hand effect proves difficult as these studies apply different methods to investigate the hot hand.

In general, the modelling framework considered provides great flexibility with regard to distributional assumptions.
In particular, the response variable is not restricted to be Bernoulli distributed (or Gaussian, as is often the case when making inference on continuous-time SSMs), such that other types of response variables used in hot hand analyses (e.g.\ Poisson) can be implemented by changing just a few lines of code. 
Our continuous-time SSM can thus easily be applied to other sports, and the measure for success does not have to be binary as considered here. 
For readers interested in adopting our code to fit their own hot hand model, the authors can provide the data and code used for the analysis.

\section*{Acknowledgements}

We thank Roland Langrock, Christian Deutscher, and Houda Yaqine for stimulating discussions and helpful comments.

\end{spacing}

\end{document}